\documentstyle[12pt,epsfig]{article}
\def\NPB{{\em Nucl. Phys.} B}
\def\PLB{{\em Phys. Lett.}  B}

\def\PRD{{\em Phys. Rev.} D}
\def\ZPC{{\em Z. Phys.} C}
\def\PRep{{\em Phys. Rep.} }
\def\M{multiplicity }
\def\MDs{multiplicity distributions }
\def\MFs{multiplicity fluctuations }
\def\FMs{factorial moments }
\def\Cu{cumulants }
\def\Co{correlations }
\def\Fq{$F_q$ }
\def\Kq{$K_q$ }
\def\Hq{$H_q$ }
\def\Pn{$P_n$ }

\def\D{distribution}
\def\Ds{distributions}

\def\be{\begin{equation}}
\def\ee{\end{equation}}
\def\bea{\begin{eqnarray}}
\def\eea{\end{eqnarray}}
\begin{document}
%%%%%%%%%%%%%%%%%%%%%%%%%%%%%%%%%%%%%%%%%%%%%%%%%%%%%%%%%%%%%%%%%%%%%%%%%%%%
%
%\begin{flushright}
%*DRAFT-00\\
%\ \ \ \ \\
%\ \ \ \ \\
%\end{flushright} 

%\title
\begin{center}
{\Large \bf What one can learn  about the QCD parton cascades studying 
the multiplicity distributions at HERA ?
}\\

\vspace*{1cm}
{ \bf Boris B. Levtchenko}\\
\vspace*{0.5cm}
{ Institute of Nuclear Physics, Moscow State University,\\
119899 Moscow, Russia }

\date{}

\vspace*{0.5cm}
 
{ ABSTRACT}\\
\end{center}
%\vspace*{0.1cm}
New properties of the multiplicity distributions 
predicted by higher order 
QCD and their physical origin are discussed briefly.
Several studies which can be performed at HERA are proposed.

\vspace*{4.5cm}
\begin{center}

Talk  presented at the \\
Joint International Workshop:\\
VIIIth Workshop on High Energy Physics and Quantum Field Theory\\
and IIIth Workshop on Physics at VLEPP\\
(Zvenigorod, Russia, September 15-21, 1993)
\footnote{e-Print version of \cite{BBL}.}
\end{center}

\newpage
%\twocolumn[

\section{\bf Definitions}

Multiplicity distributions (MD) of particles produced in high energy 
collisions are the most typical and widely discussed characteristic of
the interaction dynamics. In a condensed form MD provide information about
the fluctuations of energy spent for multiple particle production during
 a collision. 

The goal of the present paper is to review briefly the new features of the
\MDs predicted by higher order QCD.

There are two complementary ways of dealing with \MFs:

 -- studying the \D\  $P_n\,=\,\sigma_n/\sigma$ which is the number
of produced particles per event, or
 
 -- measuring the inclusive \M correlators.
 
 In practice, one uses often the normalized \FMs \Fq and \Cu \Kq 
 (for review see \cite{DW-D-K}) defined as
\be F_q\,=\,\sum^{\infty}_{n=0}n(n-1)...(n-q+1)P_n/\langle n\rangle^q
 \,=\,\frac{\langle n(n-1)...(n-q+1)\rangle}{\langle n\rangle^q}\,,
\ee

\be K_q\,=\,F_q\,-\,\sum^{q-1}_{m=1}C^{m}_{q-1}K_{q-m}F_m\,. 
\ee
Here $C^m_q\,=\,\frac{q!}{m!(q-m)!}$ are the binomial coefficients and 
$F_0\,=\,F_1\,=\,K_1\,=\,1$.

These moments have an important advantage over the original moments \cite{BP}.
The average shown in (1) implies mean value of the corresponding
expressions over the available set of experimental events. In experiment this
averaging takes into account both statistical and dynamical effects. If one 
assumes that random fluctuations  due to limited number of detected particles
are described by the Poissonian distribution, then the total average of the
 \FMs 
is equivalent to the dynamical average of usual moments \cite{BP}.

In the Feynman diagram's language, \Fq corresponds to the set of all graphs 
while the
\Cu \Kq describe the connected graphs only. The  \Cu provide the knowledge 
about
the "true" \Co, non-reducible to the product of the \Co of lower orders.
At asymptotic energies the normalized \FMs (as well as the ordinary ones)
do not depend on energy and are the functions of their rank only. The higher 
the rank of the moment is the more sensitive \Fq and \Kq are to the "tail" 
of MD
at large $n$. The  steeper decrease of the distribution at large $n$ leads to
smaller values of  the high rank \FMs.

In a theoretical analysis instead of  studying of the numerical series $P_n$ \
it is more convenient
to  analyse the function "generating" it, namely the generating
function (GF). \Fq and \Kq are easily calculated if the generating function
$G(u)$ is known \cite{DW-D-K}
\be G(u)\,=\,\sum^{\infty}_{n=0}P_n(1\,+\,u)^n\,.
\ee
Then
\be P_n\,=\,\frac{1}{n!}\frac{d^nG(u)}{du^n}\biggl |_{u=-1}\,,
\ee
\be F_q\,=\,\frac{1}{\langle n\rangle^q}\frac{d^qG(u)}{du^q}\biggl |_{u=0}\,,
\ee
\be K_q\,=\,\frac{1}{\langle n\rangle^q}\frac{d^q\ln G(u)}{du^q}\biggl
|_{u=0}\,.
\ee
Thus, the knowledge of GF gives us a possibility to calculate both the
multiplicity distribution and cumulant and factorial moments i.e. (3)-(6)
demonstrate mathematical equivalence the description of MD by functions
$P_n$, \Fq and \Kq. In \cite{Dr1} it has been proposed to use the ratio of
cumulant to \FMs $H_q\,\equiv\,K_q/F_q$ which behaves in a qualitatively
different way for various \Ds\ and is more sensitive to  specific  features
of $P_n$ which are invisible when just plotted $P_n$ or even \Fq (see Sec.\,3).

%    2

\section{\bf Some properties of the multiplicity
 distributions } 
In pre-QCD time Koba, Nielsen and Olesen published the paper \cite{KNO}
with a hypothesis about the scaling properties of the \MDs at asymptotic
energies (the KNO scaling). If $z$ is the scaled multiplicity 
$z\,=\,n/\langle n\rangle$, then the KNO scaling implies a universal form
$$\psi (z)\,=\,\langle n\rangle\,P_n$$
for the multiplicity distribution. During last 30 years the KNO-like
 behavior of
MD was experimentally confirmed in various types of high energy particle
production processes except the data on proton-antiproton interactions at
the highest energies $\sqrt{s}\,=\,$546 and 900 GeV obtained by
 UA5 collaboration
\cite{UA5} in CERN.

The negative binomial distribution (NBD)
$$G(u)\,=\,\biggl (1\,-\,\frac{u\langle n\rangle}{k}\biggr )^{-k}\,,$$
\be
P_n\,=\,\frac{(n+k-1)!}{n!(k-1)!}\biggl (\frac{\langle n\rangle/k}
{1+\langle n\rangle/k}\biggr )^{n}\biggl (1\,+\,
\frac{\langle n\rangle}{k}\biggr )^{-k}
\ee
$$F_q\,=\,\frac{(k+1)\cdot\cdot\cdot(k+q-1)}{k^{q-1}}\,,\ \ \ \ 
K_q\,=\,\frac{(q-1)!}{k^{q-1}}$$
\be
H_q\,=\,\frac{(q-1)!}{(k+1)\cdot\cdot\cdot(k+q-1)}
\ee
is another example of the distribution which is in good agreement with
experimental data
in full phase space and in smaller phase space domains. 
NBD depends on two parameters, 
the average multiplicity $\langle n\rangle$ and a positive
parameter $k$ describing the shape of the distribution.
 Here we will mention only two classes of mechanisms proposed to
generate NBD, (partial)stimulated emission \cite{Car,GV} and  cascading
\cite{GV}.

%For the later use  we point out here one feature of \Hq for NBD which follows
%from (8). Namely,  $H_q$ is ever positive and tends to zero at high ranks 
%as $q^{-k}$.
%For the Poisson distribution $H_q$ is identically equal to zero (except of 
%$H_1$\,=\,1).

One feature of \Hq for NBD is that it always positive and tends to zero
with a $q^{-k}$ behaviour at high ranks. For the Poisson distribution $H_q$ 
is identically equal to zero (except for $H_1$\,=\,1)

%  3

\section{What does QCD tell us about the \MDs?}
The KNO hypothesis was strongly supported by QCD when the equations for
generating function were solved in the so-called double logarithmic
approximation (DLA). DLA happens to be too crude however for making reasonable
predictions even for asymptotic energies: the predicted KNO shape of
the \D\ appeared to be much wider than the experimental one. On the qualitative
level, DLA can be thought to overestimate cascading processes, ignoring
completely  energy-momentum conservation since the energy of the radiating
particles remains unchanged after a soft gluon emission. Therefore 
 DLA apparently overestimates the gluon multiplicity 
because  the  parton characteristic energy  is higher  and the parton
multiplicate more actively. Taking into account  higher order perturbative
corrections  leads to a more accurate control over the parton splitting
processes and energy conservation.

Such an approach  has been realized (see \cite{BCM}, \cite{DKMT})
in the framework of the modified leading logarithmic approximation
 (MLLA) by a generalization of the standard
LLA scheme following the logic of the famous 
Dokshitzer-Gribov-Lipatov-Altarelli-Parisi approach and including the exact
angular ordering (AO) (instead of the strong AO within DLA). Thus the system of
the MLLA integro-differential equations for the quark and gluon GF has been
derived.

A recent series of publications \cite{Dr1}, \cite{YuD}--\cite{DH} was
devoted to solving of these equation in the case of $e^+e^-$-collisions with
account of different next-to-next-to leading (NNL) effects. Corresponding
corrections can be looked upon \cite{CT} as being due to a more
 accurate account of
energy conservation in the course of parton splitting. 
For example, the
approximation used in \cite{YuD}  allowed in the framework of
gluodynamics
the derivation of  
  analytical expressions for the asymptotic behaviour of \FMs
and the KNO function, which are  in better   agreement 
with the data,  by reducing substantially
the width of the theoretical distribution. Cumulant and \FMs of the \MDs in
the perturbative gluodynamics have been calculated in \cite{Dr1}, \cite{DN}.
Accounting for  the degrees of freedom  associated with  quarks
\cite{DLN} does not change the essential
qualitative features of \Fq, \Kq and  influence only weakly \Hq. 
The exact solutions of the QCD equations for quark and gluon GF 
are obtained for  the case of
fixed coupling  in \cite{DH}.

The ratio \Hq is more sensitive to the form
of \Pn at large $n$ than \Fq (see Fig. 1). It was shown in \cite{DLN}
that the predictions of \Fq shown in Fig. 1a, have qualitatively the same 
behaviour and are very close
to each other for $q\leq$\,10. However  \Hq (Fig. 1b) demonstrate much stronger
sensitivity
to the assumptions used. The most typical feature of the ratio \Hq predicted by
QCD \cite{DN} is its quasi-oscillating form with a changing  sign (Fig. 2).
Such an oscillating behavior of \Hq is a specific property of higher order QCD.
Less complete account of nonlinearities in the equation for GF leads \cite{Dr1},
\cite{DLN} only to one minimum with a very small 
value of \Hq\, (the solid line in Fig. 1b).

The results of \cite{Dr1} have initiated a search for the peculiarities of \Hq
from the experimental data. According to the \Hq measurements from \MDs
 in  $e^+e^-$--annihilation in the energy range from 22 to 91 GeV, and in
$hh$--collisions, in the energy range from 24 to 900 GeV, made in \cite{ Gia},
its behaviour corresponds to the predic-
\newpage
%%% Figs 1, 2
\vspace*{1.0cm}
\hspace{-0.5cm}
\begin{minipage}[t]{7.5cm}
\setlength{\unitlength}{\textwidth}
\begin{picture} (0.5,0.95) (0.1,0)
\mbox{\epsfig{file=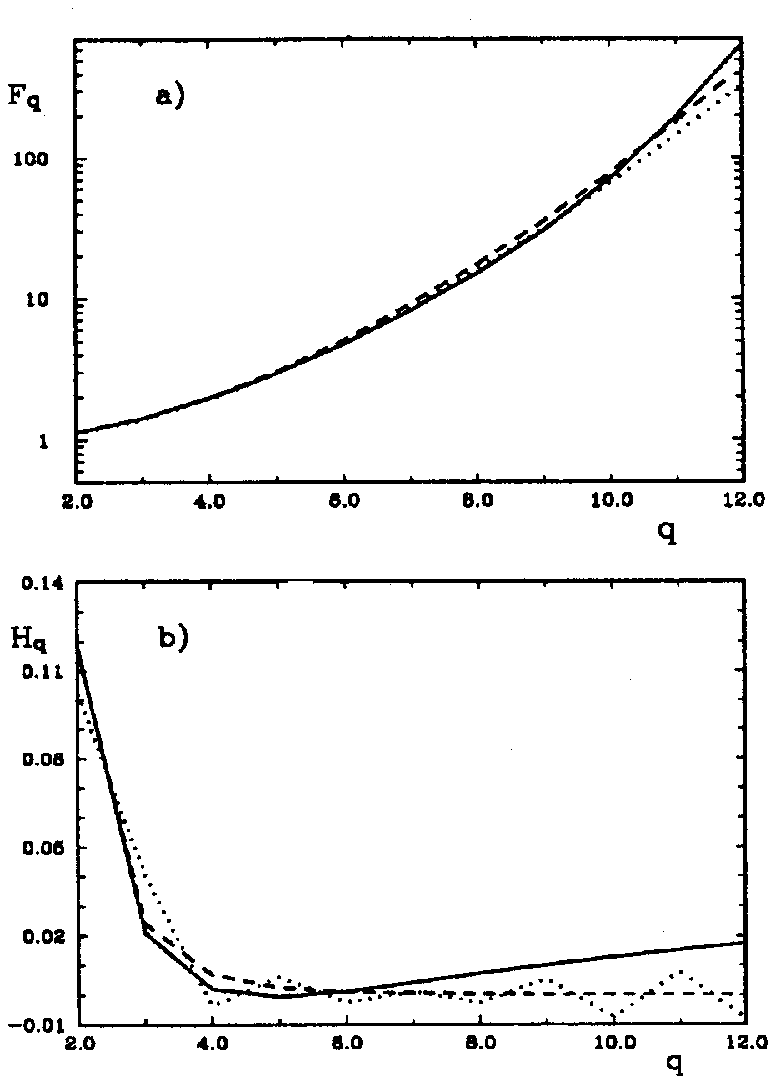,width=1.20\textwidth,height=1.35\textwidth}}
\end{picture}

 {Fig.1. a) The \FMs for different QCD \Ds\ (\cite{Dr1} - solid line;
   \cite{YuD} -- dotted line) 
 and for NBD with $k$=7.6\,(dashed line). b) The ratio \Hq 
 for the same \Ds\ as in a).}
\end{minipage}
\hspace{4mm}
\begin{minipage}[t]{7.5cm}
\setlength{\unitlength}{\textwidth}
\begin{picture} (1.0,0.95) (0.02,0)
\mbox{\epsfig{file=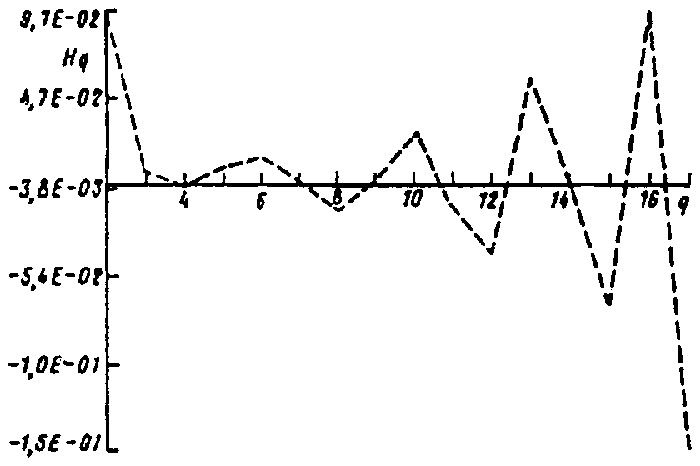,width=1.20\textwidth,height=1.10\textwidth}}
\end{picture}
%\vspace*{1.0cm}
{Fig. 2. The ratio \Hq predicted by QCD \cite{DN}.}
\end{minipage}
%%%%

\vspace*{1.0cm}
\noindent
tions of higher order QCD. A few examples are presented in Fig. 3. 
It is a surprise  for us that
 the theoretical results
 obtained for hard processes at asymptotic energies, are in a qualitative
 agreement with experimental data at low and high energies   
 both for $e^+e^-$ processes and soft hadronic collisions.

The behaviour of \Hq for NBD shown in Fig 1, where \Hq  falls monotone but
always positive,  tending to zero at large ranks $q$, is not compatible
to results shown in Fig 3.
Therefore, despite  the fact that NBD fits experimental MD very well
it is not appropriate for the complete of description 
 MD in particle production processes as claimed in \cite{DLN}
and \cite{Gia}. However, as  will be seen from Sec.5, after modifications  
NBD is able to generate the oscillating \Hq as well.

%    4

\section{ Monte Carlo Generators}

All Monte Carlo (MC) generators for high energy physics \cite{Sj1} and, in
particular, those which simulate deep inelastic scattering (DIS) \cite{HERA}
 are based on the leading logarihm (LL) picture with two body parton splitting 
 $a\rightarrow d\,+\,c$. However, as one mentioned in the previous section,
 higher orders in the perturbative QCD are necessary for a proper
  description of
 multiproduction at high energies.
 
 At present this can only be achieved in the  generators
 through approximate methods implemented in different
  QCD cascades e.g. the Lund parton shower (PS) \cite{AGIS}, 
  the color

\newpage
%%% Figs 3, 4
\vspace*{3.0cm}
\hspace{-0.5cm}
\begin{minipage}[h]{7.5cm}
\setlength{\unitlength}{\textwidth}
\begin{picture} (0.5,0.95) (0.1,0)
\mbox{\epsfig{file=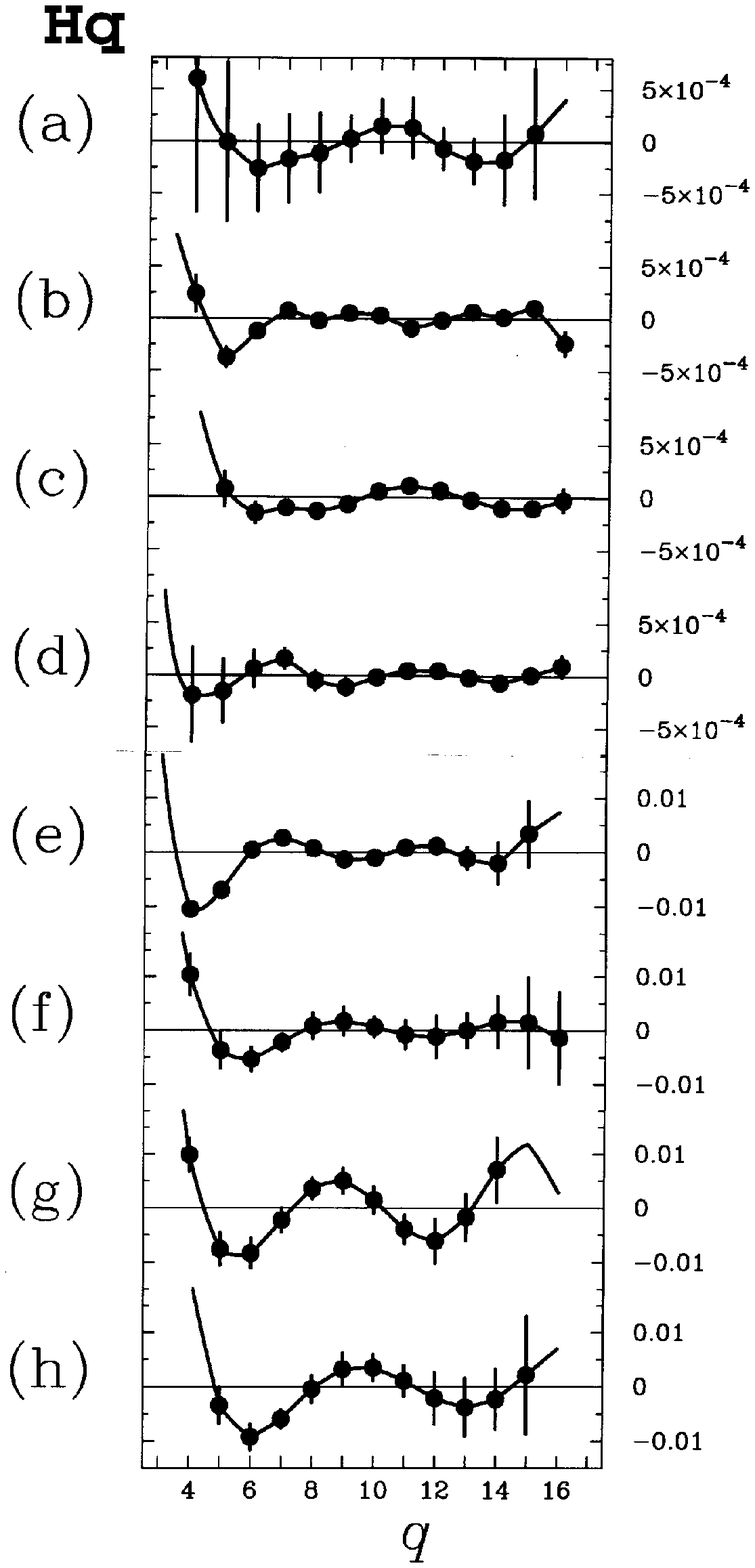,width=1.25\textwidth,height=1.65\textwidth}}
\end{picture}

 {Fig.3. Experimental data \cite{Gia} on \Hq for a)-d) $e^+e^-$ ($\sqrt{s}$=29,
 34.8, 43.8, 91 GeV) and e)-h) $hh$ ($\sqrt{s}$=62.2, 200, 546, 900 GeV)
 collisions. Lines are to guide the eye.}
\end{minipage}
\hspace{4mm}
\begin{minipage}[h]{7.5cm}
\setlength{\unitlength}{\textwidth}
\begin{picture} (1.0,0.95) (0.02,0)
\mbox{\epsfig{file=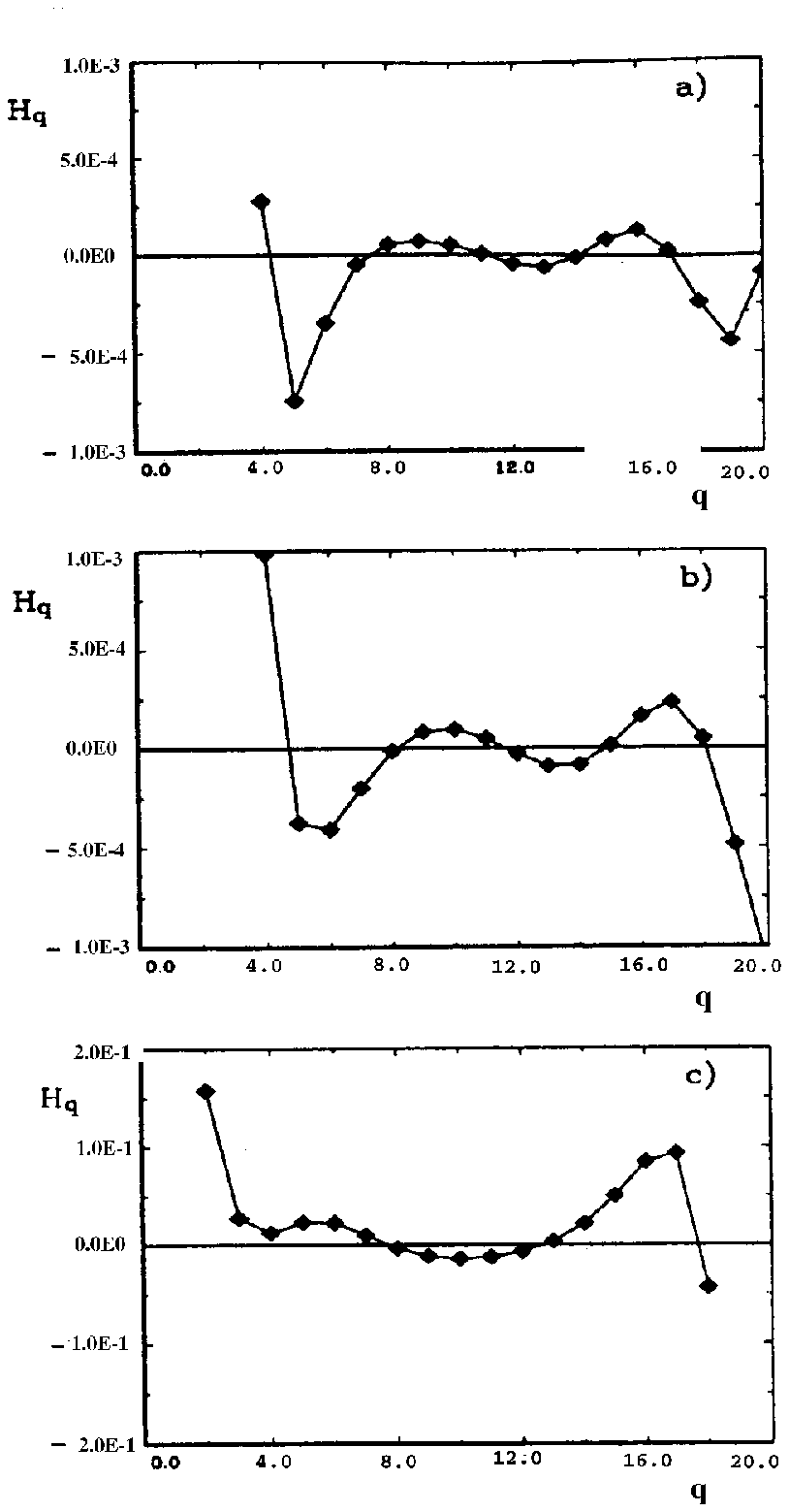,width=1.00\textwidth,height=1.50\textwidth}}
\end{picture}
%\vspace*{1.0cm}
{Fig. 4. The ratio \Hq due to the QCD MC codes:
a)  JETSET 7.3, $e^+e^-$, $\sqrt{s}$= 91 GeV; 
b) ARIADNE 4.4, $e^+e^-$, $\sqrt{s}$= 91 GeV; 
c) PYTHIA 5.5, $e^-p$, $\sqrt{s}$= 314 GeV.
Lines are to guide the eye.}
\end{minipage}
%%%%
 
\vspace*{1.0cm}
\noindent   
 dipole model (CDM)\cite{GAL}.
 The LLA used in PS and CDM does 
not  give a proper treatment of hard emissions. A method
 was developed to let a single hard emission to be controlled by the exact 
 $O(\alpha_s)$ or $O(\alpha_s^2)$ QCD matrix elements and then modelling
 subsequent radiation using the PS technique. 
    
 One can ask a question: are the above-mentioned improvements of the MC models
 enough for a proper description of \Hq\,? The answer  seems to us obvious:
 since LLA is the base of PS one should not expect an oscillatory behavior of
 \Hq. However, according to our calculations of the correlators with the MC
 generators JETSET 7.3 \cite{Sj2}, ARIADNE 4.4 \cite{HERA}($e^+e^-,
 \ \sqrt{s}\,=\,$91 GeV)
 and PYTHIA 5.5 \cite{HERA}($e^-p,\ \sqrt{s}\,=\,$314 GeV) \Hq has,
 nevertheless, an oscillating form as shown in Fig. 4.
 An explanation of such a phenomenon can be found immediately if one recall
 two facts: 1) each MC generator takes a special care about both the
   local (in the course
 of parton splitting) and global energy-momentum conservation in the collision;
 2) the finite energy of collisions  is the physical origin of large 
 $O(\alpha_s)$ corrections \cite{CT}. Thus, the LLA 
 in conjunction with  the
 energy-momentum conservation in the MC models imitate in some part the higher
 order corrections  leading to the oscillation of \Hq. 
 The question arises, though,
  how much of the higher order corrections are accounted for?

\newpage 
%%% Figs 5
%\vspace*{3.0cm}
%\hspace{-0.5cm}
\setlength{\unitlength}{\textwidth}
\begin{picture} (0.5,0.95) (0.1,0)
\mbox{\epsfig{file=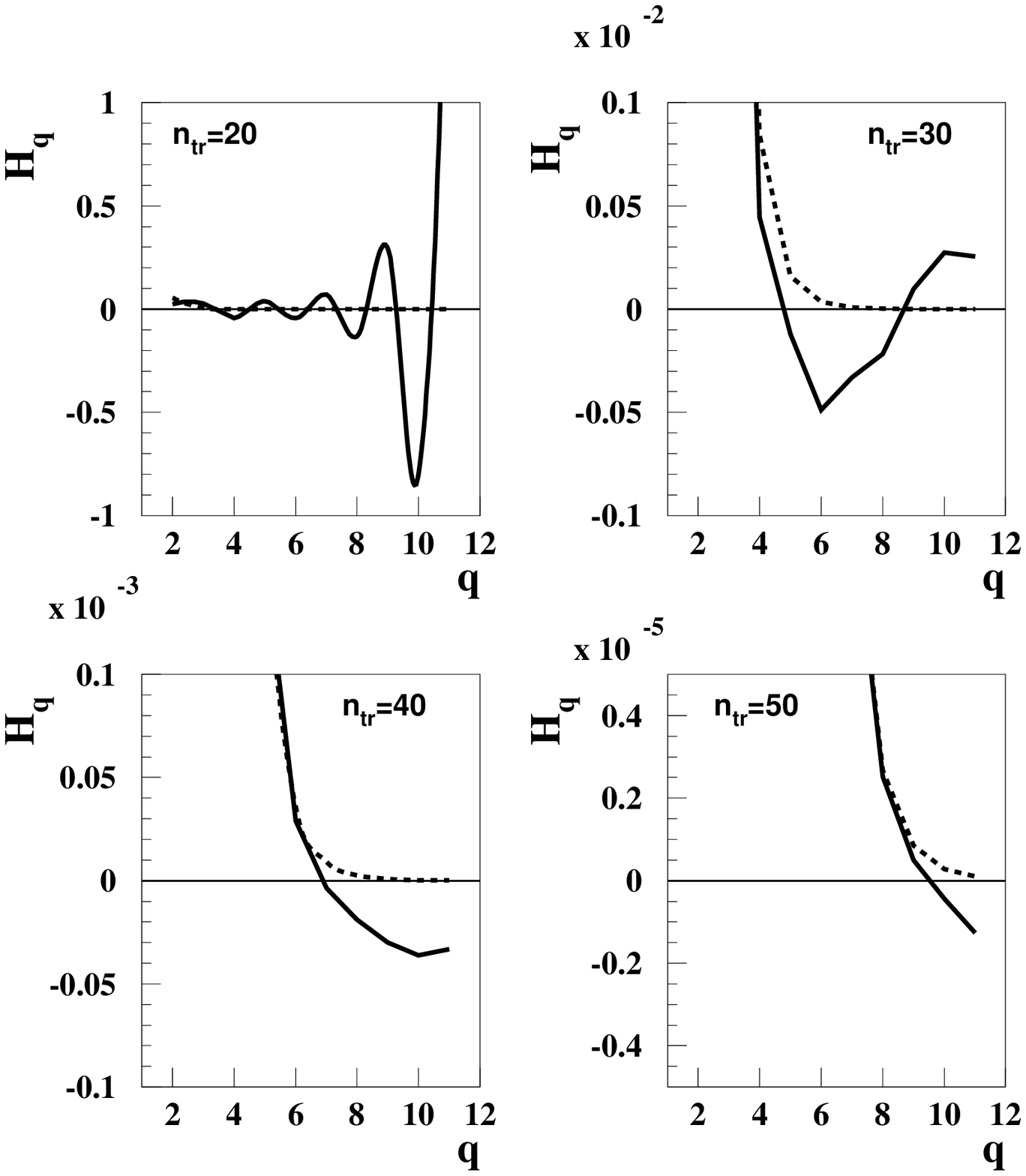,width=1.10\textwidth,%
height=1.05\textwidth}}
\end{picture}
 
 {Fig.5. The ratio \Hq calculated for NBD with $\langle n\rangle$=9.22, 
 $k$=17.24. The solid line are for NBD truncated at $n_{tr}$ and the dished 
 lines are for NBD without truncation.} 

 %5
 
 \section{Phenomenological examples}
The conclusions from the previous section can be confirmed by the following
arguments  \cite{L1}. Formally, according to (7) NBD has an infinite
 "tail" at finite  collision energy (finite $\langle n\rangle$). This
  results in positive \Kq and monotone declining \Hq (8).
 On the other hand, an infinite "tail" of MD is possible only  for
 production of massless particles or 
 through neglecting  energy conservation during the reaction. 
 Taking into account  these factors leads to a truncation of the MD
 "tail" at some finite multiplicity $n_{tr}(s)$. As a result,
  \Hq calculated for the
 truncated NBD  oscillates around the curve $q^{-k}$ with alternating 
 sign. The amplitude of the oscillation tends  to zero quickly
 as  $n_{tr}\rightarrow\infty$ 
 and $H_q^{(tr)}$ tends to $H_q^{(NBD)}$ (Fig. 5). The same behaviour of 
 \Hq has been found for the truncated Poisson distribution (PD).
 
 Another example of the behaviour of \Hq is in soft $p\bar{p}$
  collisions at $Sp\bar{p}S$
 and Tevatron energies calculated in \cite{LS} in the framework of the Dual
 Parton Model \cite{CTTV}. It was found \cite{LS} the properties of \Hq 
 (  amplitude of the oscillation, positions of minima and maxima) are very
 sensitive to the number of cutted Pomerons accounted for in the calculation.
 
 %   6
 \section{ What can be done at HERA\,?}
 In  high energy reactions used in the study of  the oscillations of \Hq  
 \cite{Gia} only DIS data are   missing. 
 New data from the $ep$ collider HERA
 will be able to rectify  this situation. The invariant mass $W$ of
  the hadronic final
 state in DIS at HERA extends, with significant cross sections, to the phase
 space limit ($\sqrt{s}\,=\,$314 GeV). This circumstance allows us to formulate
 several problems related to properties of MD which can be studied with the H1
 and ZEUS detectors:
 
 1. Detailed study of MD as a function of $z\,=\,n/\langle n\rangle$ over the
 whole kinematical region of $W$. Does the KNO
 scaling  violated at large W ?
 
 2. High precision measurement of the ration $H_q\,=\,K_q/F_q$  of cumulant
 and factorial moments both for the full phase space and restricted rapidity
 windows, for events with 1+1, 1+2, ... jets, etc. Does \Hq as
  a function of the
 order $q$ shows an oscillations around $H_q\,=\,0$\,? If so, confronting
 the data with predictions of the MC models we would  learn more about 
  the higher order effects implemented in these  MC models.
 
 3. Measurements of \Hq at different $W$   will shed more light on the problem
 how the finite energy effects influence the \Hq shape.
 
 To conclude,  energy-momentum conservation plays a very important role
 in the correct description of the \MDs, \Fq, \Kq and \Hq in the
 framework of QCD and different phenomenological models. The ratio \Hq 
 is extremely sensitive to the length  of the MD "tail".
  In  perturbative QCD the behaviour of the MD "tail" is controlled by
  higher order corrections while for   phenomenological approaches (NBD, PD etc.)
  the finite energy effects have to be accounted for 
  by  truncating  the MD "tail".

\vskip 0.4cm
\noindent{\bf Acknowlegments.} The author thank I.V. Andreev, I.M. Dremin, 
and G. Gianini for discussions, N. Brook for reading the manuscript
 and comments.
This work was supported in part by  the
International Science Foundation under grant  Ph1 35\,08045 and DESY.

%\newpage

%
%%%%%%%%%%%%%%%%%%%%%%%%%%%%%%%%%%%%%%%%%%%%%%%%%%%%%%%%%%%%%%%%%%%%%%  

\begin{thebibliography}{99}
\bibitem{BBL}
B.B.\,Levtchenko, in {\it Proc.of the  Joint Int. Workshop: VIII QFTHEP 
and III VLEPP Physics Workshops},(Zvenigorod, Russia, Sept. 15-21, 1993). 
Ed. B.B.\,Levtchenko,
MSU-Press, Moscow, 1994, p.\,68.
\bibitem{DW-D-K}
E.A.DeWolf, I.M.Dremin and W.Kittel, Usp. Fiz.Nauk 163\,(1993)\,3.
\bibitem{BP} A.Bialas, R. Peschanski, \NPB\,273\,(1986)\,703.
\bibitem{Dr1}
I.M. Dremin, \PLB\,313\,(1993)\,209.
\bibitem{KNO}
Z.\,Koba, H.B.\,Nielsen and P.\,Olesen, \NPB\,40\,(1972)\,317.
\bibitem{UA5}
G.J.\,Alner et al.\,(UA5), \PRep 154\,(1987)\,247.
\bibitem{Car}
P.\,Carruthers, C.C.\,Shin, \PLB 127\,(1983)\,242.
\bibitem{GV}
A.\,Giovannini and L.\,Van Hove, \ZPC\,30\,(1986)\,391.
\bibitem{BCM}
A.\,Basseto, M.\,Ciafoloni and G.\,Marchesini, \PRep\,100\,(1983)\,202.
\bibitem{DKMT}
Yu.L.\,Dokshitzer, V.A.\,Khoze, A.H.\,Mueller and S.I.\,Troyan,
{\em Basics of Perturbative QCD}. Ed. J. Tran Thanh Van,
 Editions Frontieres, 1991.
\bibitem{YuD}
Yu.L.\,Dokshitzer, \PLB\,305\,(1993)\,295.
\bibitem{DN}
I.M.\,Dremin and V.A.\,Nechitailo, JETP Lett. 58\,(1993)\,881.
\bibitem{DLN}
I.M.\,Dremin, B.B.\,Levtchenko and V.A.\,Nechitailo
{\em Phys.Atom.Nucl.}\,57\,(1994)\,1029. 
\bibitem{DH}
I.M.\,Dremin and R.C.\,Hwa,\PRD\,49\,(1994)\,5805
 \PLB 324\,(1994)\,477.
\bibitem{CT}
F.\,Cuypers and K.\,Teshima, \ZPC\,54\,(1992)\,87.
\bibitem{Gia}
G.\,Gianini et al., in:{\it Proc. XXIII Inter.
 Symposium on Multiparticle Dynamics},
 (Aspen, USA, Sep.12-17, 1993), Eds M.M.\,Block, A.R.\,White, World Scientific,
  Singapore, 1993, p. 405.
\bibitem{Sj1}
T.\,Sj\"{o}strand, Monte Carlo Event Generation for LHC,
 Preprint CERN-TH.6275/91.
\bibitem{HERA}
Physics at HERA, {\it Proc. of the Workshop}, Eds. W. Buchm\"{u}ller, 
G. Ingelman, Hamburg, 1991, v.\,3.
\bibitem{AGIS}
B.\,Andersson, G.\,Gustafson,  G.\,Ingelman and T.\,Sj\"{o}strand,
 Phys.Rep. {\bf 97}\,(1983)\,31.
\bibitem{GAL}
G.\,Gustafson, \PLB 175\,(1986)\,453; 
B.\,Andersson, G.\,Gustafson and  L.\,L\"{o}nnblad, \NPB 339\,(1990)\,393. 
\bibitem{Sj2}
T.\,Sj\"{o}strand, CERN.TH.7112/93.
\bibitem{L1}
B.B.\,Levtchenko, to be published.
\bibitem{LS}
B.B.\,Levtchenko and A.V.Shumilin, to be published.
\bibitem{CTTV}
A. Capella, J. Tran Thanh Van, \ZPC\,23\,(1984)\,165.

\end{thebibliography}
\end{document}